\documentclass[prl,showpacs,superscriptaddress,twocolumn]{revtex4-2}
\usepackage{amssymb,amsmath,xcolor,graphicx,psfrag}

\usepackage{hyperref, cancel}

\begin{document}

\newcommand{\jav}[1]{#1}
\newcommand{\rim}[1]{{\color{green}#1}}
\newcommand{\rin}[1]{{\color{blue}#1}}

\title{Adiabaticity enhancement in the classical transverse field Ising chain,\\ and its effective non-Hermitian description}

\author{Bal\'azs D\'ora}
\email{dora.balazs@ttk.bme.hu}
\affiliation{MTA-BME Lend\"ulet Topology and Correlation Research Group,
Budapest University of Technology and Economics, M\H uegyetem rkp. 3., H-1111 Budapest, Hungary}
\affiliation{Department of Theoretical Physics, Institute of Physics, Budapest University of Technology and Economics, M\H uegyetem rkp. 3., H-1111 
Budapest, Hungary}
\author{Roderich Moessner}
\affiliation{Max-Planck-Institut f\"ur Physik komplexer Systeme, 01187 Dresden, Germany}

\date{\today}

\begin{abstract}
We analyse the near-adiabatic dynamics in a ramp through the critical point (CP) 
of the {\it classical} transverse field Ising chain. This is motivated, conceptually, by the fact that this CP--unlike its quantum counterpart--experiences no  thermal or quantum fluctuations, and technically by the tractability of its effective model. 
For a `half-ramp' from ferromagnet to CP, the longitudinal and transverse magnetization scale as $\tau^{-1/3}$ and $\tau^{-2/3}$, respectively, with $1/\tau$
the ramp rate, in accord with Kibble-Zurek theory.
For ferro- to paramagnetic ramps across the CP, however, they  stay {\it closer}, $\tau^{-1/2}$ and $\tau^{-1}$, to adiabaticity. 
This  adiabaticity enhancement  compared to the half  ramp is understood by casting the dynamics in the paramagnet in the form of  a non-hermitian Dirac {\it Hamiltonian}, with the CP playing the 
role
of an exceptional point, opening an additional decay channel. 
\end{abstract}

\maketitle

\paragraph{Introduction.}
The study of the non-equilibrium dynamics of many-body systems has  attracted enormous interest 
recently\cite{polkovnikovrmp,dziarmagareview}. In particular, defect production after the non-adiabatic 
passage through a critical
point was found to exhibit universal behavior with a scaling that is determined solely by the universality class 
of the underlying phase transition\cite{kibble,zurek}. This, the celebrated Kibble-Zurek mechanism, has been most directly 
verified in the transverse field Ising chain\cite{zurekprl,dziarmaga}.

Indeed, the transverse field quantum Ising chain~\cite{Elliott_TFIM_PRL} is a paradigmatic, and one of the most thoroughly studied, models in physics.
It plays an important role in understanding quantum phase transitions\cite{sachdev} and duality, conformal field theory\cite{mussardo} 
and is relevant, through a quantum to classical mapping,  for the statistical mechanics of the classical Ising model in two dimensions. 
Moreover, it is closely tied to topological phenomena\cite{laumann} and the Kitaev chain\cite{kitaev2001}. In addition to its theoretical 
appeal, experimental realizations involve
condensed\cite{bitko,kinross} and artificial\cite{keesling} matter.

Surprisingly, despite the great amount of interest in the transverse field quantum Ising chain, its classical counterpart has 
received only limited attention. This is all the more remarkable as the study of classical spin chains has unearthed a number of surprises~\cite{gerling,schubert}, such as a regime of Kardar-Parisi-Zhang scaling in the Heisenberg chain~\cite{Heisenberg_KPZ} or generalized hydrodynamics\cite{doyon}. Moreover, classical spin models are ubiquitous in that they are widely used not only in physics\cite{blundell,nowak,sachdev}, but also to model complex systems such as neural or social 
networks\cite{rojas}, and physical realizations include large interacting quantum spins\cite{nowak}, polariton simulators\cite{berloff}, coupled nanolaser
lattices\cite{parto}, cold atoms with total collective 
spin dynamics\cite{hoang} or precessing rigid bodies such as a spinning top \cite{opatrny}.

The identical ground state phase diagrams\cite{cuccoli,caramico,sachdev} of classical and quantum transverse field Ising chains, being non-integrable\cite{johann} and integrable\cite{sachdev}, respectively, possess 
ferromagnetic (FM) and paramagnetic (PM)
phases, separated by a CP. However,  critical exponents and the ensuing universality classes are   distinct.
Moreover, the classical  version lacks both quantum and thermal 
fluctuations, rendering the corresponding CP  unusual.

Here we investigate the non-equilibrium dynamics in the classical transverse field Ising chain for a wide variety  of ramps, starting 
from
a ground state.  
We summarise the scaling of physical quantities with the  ramp speed  in Table 
\ref{table1}. While many decay exponents are in accord with Kibble-Zurek scaling, those of deviations of the spins from adiabaticity for a ramp  from the FM
across
the CP differ--{\it they stay closer to the adiabatic limit!} Our analysis accounts for this by interpreting the classical dynamics as a non-Hermitian Dirac Hamiltonian, where the CP plays the role of an exceptional point, which allows for an additional decay channel due to effective non-unitary dynamics.

\begin{table}[h!]
\begin{tabular}{c|c|c|c|c}
\hline
ramp type \textbackslash  observables & $\delta^x(\tau)$ & $\delta^y(\tau)$ & $\delta^z(\tau)$ & $\Delta E(\tau)/N$ \\
\hline
\hline
FM $\rightarrow$ FM & 1  & 1 & 1 & 2 \\
\hline
FM $\rightarrow$ CP & 1/3  & 2/3 & 2/3 & 4/3 \\
\hline
FM $\rightarrow$ PM & {\color{blue} 1/2 } &  {\color{blue} 1/2} &  {\color{blue}1} & 1\\
\hline
\end{tabular}
\caption{The exponent of deviations from the adiabatic value for several physical quantities as $\tau^{-a}$  for various ramps. For PM to FM ramp, no defects 
are 
created. The PM initial state ${\bf S}_n=(0,0,1)$ 
remains the solution of the Landau-Lifshitz equation for any time dependent $g(t)$. The FM to FM ramp is unrelated to Kibble-Zurek theory since no 
CP is crossed. 
The scaling of the excess energy as well as the FM to CP ramp follow the Kibble-Zurek theory. The exponents in blue follow from an effective non-hermitian dynamics.}
\label{table1}
\end{table}

We study the one dimensional Hamiltonian  
\begin{gather}
H=\sum_n -JS_n^xS_{n+1}^x-2gS_n^z
\label{classham}
\end{gather}
with  unit length classical spins ${\bf S}_n \in S^2$ and ferromagnetic Ising coupling $J=1$, and transverse field strength $g>0$. We use periodic boundary conditions for the $N$ spins.
For $g<1$, the ground state is FM\cite{botet,cuccoli}
and $S_{n,gs}^x=\pm\sqrt{1-g^2}$, $S_n^z=g$, the $\pm$ signs corresponds to the two degenerate ground state configurations. The ground state energy per 
spin
is $E_{gs}/N=-2g-(1-g)^2$.
In the PM, $S_{n,gs}^z=1$ for $g>1$ with ground state energy $E_{gs}/N=-2g$.
These two regions are separated by a classical CP at $g=1$, which corresponds to a continuous 
second order classical phase transition upon tuning the transverse field with critical exponents 
$\alpha=0$, $\beta=1/2$, without thermal or quantum fluctuations at zero temperature. 
The non-analytic behaviour of the transverse magnetization at $g=1$ also allows us to define the corresponding critical exponent $\beta_z=1$.

The spin dynamics of the model is obtained from the classical Landau-Lifshitz equation of motion
\begin{gather}
\partial_t {\bf S}_n=
{\bf B}_n\times {\bf S}_n,
\end{gather}
where the effective magnetic field for the $n$th spin is
\begin{gather}
{\bf B}_n=-(S_{n-1}^x+S_{n+1}^x,0,2g).
\end{gather}
A classical linear spin wave theory\cite{blundell} analysis reveals that in the PM, the energy spectrum is $\omega_q=2\sqrt{g(g-\cos(q))}$ with $q$ 
the momentum. 
The energy
disperses linearly with momentum around $q=0$ at the CP $g=1$,  giving the dynamical critical exponent $z=1$, while the gap collapses as
$\sim\sqrt{g-1}$ upon approaching the critical point\cite{botet}, defining the exponent of the correlation length as $\nu=1/2$.
In the FM, the spin wave spectrum is $\omega_q=2\sqrt{1-g^2\cos(q)}$, yielding the same $\nu$ and $z$.
During any non-equilibrium dynamics, the system heats up, and thermal fluctuations appear.

Similarly to its quantum counterpart\cite{zurekprl,dziarmaga,dziarmagareview,polkovnikovrmp}, we are interested in ramping the transverse field
as $g(t)=g_0+(g-g_0)t/\tau$ across the classical CP with a speed of $1/\tau$ and $0<t<\tau$.
Since these ramps are spatially homogeneous and start from the ground state, where all spins behave identically, the dynamics involves
only  the 
homogeneous
long wavelength ($q=0$) mode of the spins, and it suffices to study the dynamics of a single spin as
\begin{gather}
\partial_t {\bf S}={\bf B}\times {\bf S} \textmd{ with } {\bf B}=-2(S^x,0,g).
\label{effll}
\end{gather}
This corresponds to the effective Hamiltonian\cite{botet}
\begin{gather}
H= -\left(S^x\right)^2-2gS^z
\label{lmg}
\end{gather}
of a single classical spin with uniaxial anisotropy  in the presence of transverse field.
Note that this applies to spatially homogeneous quenches in the ground states of the classical model in Eq. \eqref{classham}.
The spatial correlation length is infinite in the original model and loses its meaning in the effective description, where only temporal
fluctuations appear. This simplifies the problem immensely since 
only three, rather than $3N$, coupled differential equations need to be solved.
Dynamics in the quantum version of Eqs. \eqref{classham} and \eqref{lmg} was considered in Refs. 
\onlinecite{zurekprl,dziarmaga,kolodrubetz,defenu,xue2018,acevedo,caneva}.

\paragraph{Ramp from PM.} For quenches with $g_0>1$ and any final $g$, the initial spin configuration, ${\bf S}=(0,0,1)$,  
is {stationary} and remains a solution of Eq.~\eqref{effll} for any
 time dependent transverse field, in contrast to the Kibble-Zurek 
mechanism\cite{kibble,zurek,Gubser_KZ}, where the final energy density depends on the ramp rate. Here, it is determined by the energy of the infinitely long lived 'scar state'.
Loosely speaking, the system can neither choose any spin configuration (up or down in the $x$ direction) on the FM side 
for a spatially homogeneous quench, nor can it be in a superposition  due to the classical nature of the spins.
The same  applies to ramps starting from the ground state at the CP. 

\paragraph{Ramp within the FM phase.}
For simplicity, we consider $g(t)=gt/\tau$ with $0<t<\tau$ and $g<1$.
By introducing the difference
\begin{gather}
\delta^{j}(t)=S^{j}(t)-S_{gs}^{j}(g(t)) \hspace*{6mm}j=x,y,z 
\end{gather}
between the time evolved and adiabatic ground state values of the spins,
the effective dynamics to lowest order in $\delta$ is described from Eq. \eqref{effll} by only two coupled equations as
\begin{subequations}
\begin{gather}
\partial_t\delta^y(t)=2\sqrt{1-(gt/\tau)^2}\delta^z(t),\\
\partial_t\delta_z(t)=-2\sqrt{1-(gt/\tau)^2}\delta_y(t)-\frac{g}{\tau} \label{deltaz1}
\end{gather}
\label{ferroramp}
\end{subequations}
\noindent with initial condition $\delta^{y,z}(t=0)=0$
and $S^x(t)$ implicit via the unit length constraint. 
For $g\ll 1$ and long enough quenches, this is solved to a good approximation by taking $\sqrt{1-(gt/\tau)^2}$ as a time independent constant to yield
\begin{gather}
\delta^z(t)=-\frac{g\sin(2\int_0^t\sqrt{1-(gt'/\tau)^2}dt')}{2\tau\sqrt{1-(gt/\tau)^2}}
\label{dz1}
\end{gather}
and $\delta^y(t)$ follows from using Eq. \eqref{deltaz1}. Notably, both contain the $1/\tau$ term from the denominator of Eq. \eqref{dz1}, which stems
from the source term in Eq. \eqref{deltaz1}.
At the end of the quench $t=\tau$, all spin components deviate from their ground state
values  $\sim\tau^{-1}$.
From this and Eq. \eqref{lmg}, the naive expectation for the scaling of the excess energy would be the same. However, in reality, the excess energy 
vanishes as $\tau^{-2}$ since the $\tau^{-1}$ prefactors in the two terms in Eq. \eqref{lmg} cancel.  This is expected to be the  typical 
scaling for the near-adiabatic dynamics 
away from the CP\cite{puebla2015}.

\paragraph{Ramp from FM to CP}
In this case, we consider $g(t)=t/\tau$ with $0<t<\tau$. The dynamics is still described by Eq. \eqref{ferroramp} with $g=1$, and the time evolution ends
at the CP. From a numerical analysis of the problem, we learn that
to a good approximation, $\sqrt{1-(S^x(t))^2}\approx S^z(t)=\frac{t}{\tau}+\delta^z(t)$.


\begin{figure}[t!]
\centering
\includegraphics[width=7cm]{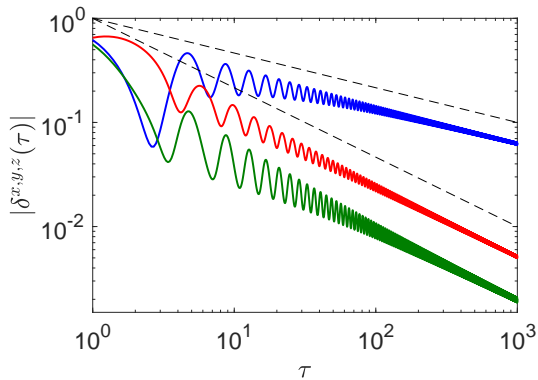}
\caption{Numerical data for deviations from the adiabatic value for spin components $x$, $y$ and $z$ (blue, red and green) from top to bottom after quenching to the critical
point ($g=1$) from the FM ($g_0=0$) phase.
The black dashed lines depict the $\tau^{-1/3}$ and $\tau^{-2/3}$ scalings.}
\label{tfifc} 
\end{figure}

At the heart of the problem lies the square root vanishing longitudinal spin component, $\sqrt{1-(t/\tau)^2}$ at the CP, which renders
any adiabatic approximation rather difficult, since any temporal derivative of the above expression diverges at the CP.
This can be cured by retaining higher order terms in $\delta$ in Eq. \eqref{ferroramp}.
In this case, we have to replace the $\sqrt{1-(t/\tau)^2}$ term by $\sqrt{1-(t/\tau+\delta^z(t))^2}$ which cures the aforementioned singularity
at the CP for any finite $\delta^z(\tau)$.
Eq. \eqref{dz1} is replaced by
\begin{gather}
\delta^z(t)=-\frac{\sin(2\phi(t))}{2\tau\sqrt{1-(t/\tau+\delta^z(t))^2}}
\label{dz2}   
\end{gather}
with $\phi(t)=\int_0^t \sqrt{1-(t'/\tau+\delta^z(t'))^2}dt'$.  The self consistency condition at the end of the quench at the CP reads
\begin{gather}
\delta^z(\tau)\approx -\frac{\sin(2\phi(\tau))}{2\tau\sqrt{-2\delta^z(\tau)}},
\end{gather}
which yields $\delta^z(\tau)\sim \tau^{-2/3}$ and $\delta^y(\tau)$ follows the same scaling via 
Eq. 
\eqref{deltaz1}.
From the unit length constraint, we obtain $\delta^x(\tau)\sim \tau^{-1/3}$.
These features are shown in Fig. \ref{tfifc} from the full numerical solution of Eq. \eqref{effll} using a 5th order Runge-Kutta method.

\begin{figure}[h!]
\centering  
\includegraphics[width=7cm]{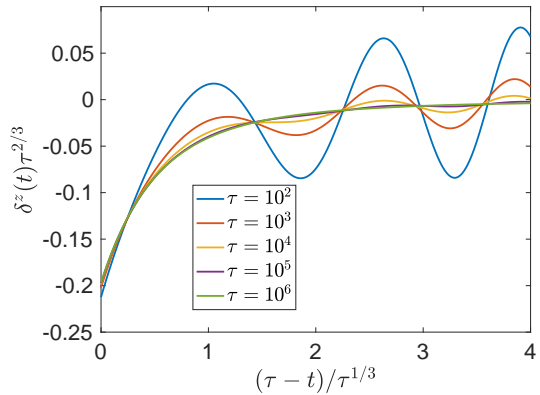}
\caption{Temporal data collapse of $\delta^z(t)$ close to the CP for various ramp rates after a ramp from the FM ($g_0=0$). The data is obtained from the numerical solution of Eq. \eqref{effll}
Deviations from zero start to appear at times $\sim\tau^{1/3}$ before reaching the CP. 
~Similar data collapse characterizes the other spin
components.}
\label{tficritical}
\end{figure}

The transition time $t_{tr}$ when this scaling appears due to the close vicinity of the CP, is determined from 
$S^x(\tau-t_{tr})\approx 0$. 
Using the above 
scalings, this gives $t_{tr}\sim \tau^{1/3}$: the presence of criticality makes its presence felt at $t\sim\tau-\tau^{1/3}$. These features are 
illustrated in Fig. \ref{tficritical}.
From the above scalings and the structure of the effective Hamiltonian in Eq. \eqref{lmg}, it is tempting to conclude that
the excess energy pumped into the system by the non-equilibrium ramp scales as $\Delta E/N\sim \tau^{-2/3}$, since both terms
in Eq. \eqref{lmg} produce this scaling. This is, however, again not correct. The prefactors of the leading order terms cancel, leaving the subleading term
for the excess energy, $\Delta E/N\sim \tau^{-4/3}$.

These exponents follow  Kibble-Zurek scaling\cite{kibble,zurek}. In general, an operator is expected to scale 
with the speed of the drive after the ramp as\cite{xue2018,chandranprb,kolodrubetz,adpert}
${\mathcal O}-{\mathcal O}_0\sim \tau^{-\chi/(1+\mu)}$ measured from its adiabatic value ${\mathcal O}_0$ with $\chi$ the 
critical exponent associated to $\mathcal O$ and $\mu=z\nu$ in  conventional Kibble-Zurek theory. 
Using the critical exponents for Eq. \eqref{classham} for the spins and  linear spin wave theory, 
we obtain $\mu=1/2$, the 
conventional value 
expected in related models\cite{defenu,xue2018}. Also the transition time\cite{dziarmaga} is expected to scale as $\tau^{\mu/(1+\mu)}$, in accord with Fig.
\ref{tficritical}.
As to the excess energy, its critical exponent is\cite{cardy} $2-\alpha=2$, which gives the observed $-4/3$ exponent. Note that these differ  
from those identified for the quantum counterparts\cite{dziarmaga,defenu,xue2018} of 
Eqs. \eqref{classham} and \eqref{lmg}.

\paragraph{Ramp from FM to PM.}
This is the continuation of the previous FM to CP ramp.
The temporal variation of the transverse field is $g(t)=gt/\tau$ with $g>1$, $0<t<\tau$. 
For the initial FM part of the ramp for $t<\tau/g$, we use the results in the previous section and
 pick up the time evolution at the critical point, $t=\tau/g$. 

By linearizing Eq. \eqref{effll}, the effective dynamics couples only the $x$ and $y$ components, resulting in two coupled differential equations.
These {\it equations of motion} may be written in a  suggestive form of a non-Hermitian {\it Hamiltonian}, namely a Schr\"odinger 
equation\cite{gao2015,rotter,zeuner,Feng2014,hodaei,Bergholtz2021,ashidareview,ElGanainy2018,fruchart}  of a quantum 
spin-1/2 in a {complex} magnetic field or the non-hermitian $(0+1)$ 
dimensional Dirac equation\cite{boozer}
\begin{gather}
i\partial_t\left(\begin{array}{c}
\delta^x(t)\\
\delta^y(t)
\end{array}\right)=\left(\begin{array}{cc}
0 & 2ig(t) \\
-2i g(t)+2i & 0
\end{array}\right)
\left(\begin{array}{c}
\delta^x(t)\\
\delta^y(t)
\end{array}\right)
\label{pararamp}
\end{gather}
and $\delta^z(t)$ follows from the unit length spin constraint,  $\tau/g<t<\tau$. 
Eq. \ref{pararamp}  is a non-hermitian, PT-symmetric\cite{mostafazadeh2002,mostafazadeh2003,Bender2007} Dirac-equation, where 
non-hermiticity arises from the $2 i \delta^x(t)$ term in Eq. 
\eqref{pararamp}. The instantaneous eigenvalues
are $\pm 2 \sqrt{g(t)(g(t)-1)}$, These vanish exactly at the CP $g(t)=1$, $t=\tau/g$, where the effective dynamics starts. This corresponds to an exceptional point\cite{heiss,ashidareview,Bergholtz2021} with vanishing 
spectrum 
 at $g=1$, consistent with critical exponent $\mu=1/2$.
There, not only the eigenvalues become degenerate but also the two \emph{eigenstates} coalesce and no longer form a complete basis. The system
becomes increasingly hermitian with time for $g\gg 1$. 
The initial condition to Eq. \eqref{pararamp} is the one and only eigenstate of the r.h.s. of Eq. \eqref{pararamp} at the exceptional point, 
namely $(\delta^x(\tau/g),\delta^y(\tau/g))\sim \tau^{-1/3}(1,0)$\footnote{In principle,
the initial condition is $\delta^x(\tau/g)\sim\tau^{-1/3}$ and
$\delta^y(\tau/g)\sim\tau^{-2/3}\rightarrow 0$ where
the latter is taken to zero as it is parametrically  smaller than the
former in the adiabatic, $\tau\rightarrow\infty$ limit.}.

This represents a variation of the theme of non-hermitian Kibble-Zurek scaling\cite{dorakz,PRXQ}. By starting the time evolution from the single 
eigenstate of the Dirac 
equation in Eq. \eqref{pararamp} at the exceptional point,
the norm of the "wavefunction", $(\delta^x(t),\delta^y(t))$ decays in time\cite{graefe2008}. This occurs even in PT-symmetric systems due to the non-unitary non-eigenstate evolution.  By taking a fixed $g(t)=g\gtrsim 1$ in Eq. \eqref{pararamp}, the norm of the initial state $(1, 0)$  evolves in time as $1-g^{-1}\sin^2(Et)$ with $E=2\sqrt{g(g-1)}$, which decays initially before revival sets is.  However, by reintroducing $g(t)$, when the driving rate $\partial_t E/E$ is larger than the revival frequency $E$, there is not enough time for revival and only the norm decay remains.
Most of the decay occurs at the close vicinity\cite{dorakz} of the exceptional point, i.e. within a $\tau^{1/3}$ temporal 
window,
similarly
to the FM side of the transition (see Fig. \ref{tficritical}). The resulting suppression of $(\delta^x(\tau))^2+(\delta^y(\tau))^2$ from its initial value scales as $\tau^{-1/3}$. This comes from $1/3=\mu/(1+\mu)$. 
Altogether, the overall decay exponent of $\delta^x(\tau)$, {including the initial value}, is $1/2=1/3+1/6$. This is a combination of two factors: 
the $1/3$ comes from the $\tau^{-1/3}$
scaling of the initial condition, while the additional suppression factor
$1/6$   from the non-hermitian time evolution around the exceptional point, i.e. $\frac 12 \mu/(1+\mu)$.

\begin{figure}[t!]
\centering
\includegraphics[width=7cm]{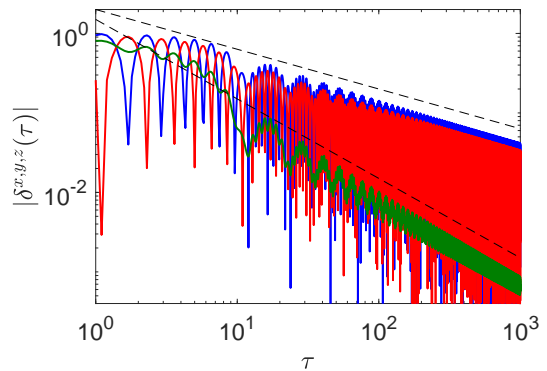}
\caption{Numerical solution of the deviations from the adiabatic value for spin components $x$, $y$ and $z$ (blue, red and green) from top to bottom after quenching from FM 
($g_0=0$) to PM ($g=3$).
The black dashed lines depict the $\tau^{-1/2}$ and $\tau^{-1}$ scalings obtained analytically.}
\label{tfifp}
\end{figure}

In addition to these scaling ideas, we treat these two coupled first order differential equations with the WKB method, similarly to Dirac systems\cite{orden}.
By solving the wavefunction away from the exceptional point, corresponding  to the 
turning point in WKB approaches, we obtain the asymptotic form of the wavefunction.
We then match this form with the initial exact solution of the linearized version of 
Eq. \eqref{pararamp}. This amounts to replacing $g(t)$ by $g$ in the first line of Eq. \eqref{pararamp}
 while keeping $g(t)=gt/\tau$ in the second line.
This is solved exactly using Airy functions and we 
obtain $\delta^x(\tau)\sim\delta^y(\tau)\sim  \exp(\pm i\tau 2\int_0^{g-1} \sqrt{t'(t'+1)}dt') ~\tau^{-1/2}$. 
The numerical solution of Eq. \eqref{effll} using the Runge-Kutta method for the FM to PM ramp is  illustrated in Fig. \ref{tfifp}.

The deviation of the $z$ component from the adiabatic value follows from the unit length constraint as 
$\delta^z(\tau)\sim (\delta^x(\tau))^2+(\delta^y(\tau))^2 \sim \tau^{-1}$, i.e. the initial value $\sim \tau^{-2/3}$ is further suppressed by
$\tau^{- \mu/(1+\mu)}$. Building on these, the excess energy also scales as $\tau^{-1}$.
 This is connected to Kibble-Zurek ideas\cite{fei}: the difference between the $1/\tau$ exponent of excess energy for a half ramp and a 
full ramp is exactly
$\mu/(1+\mu)=1/3$, which translates to $4/3-1/3=1$ in our case.

Overall, the deviations from the adiabatic value of the spin are  suppressed for a full FM $\rightarrow$ PM ramp compared to  a half ramp. This chimes with the robustness of the PM ground state spin configuration, which remains immune to any quenches.
But it is in contrast to the response of the quantum transverse field Ising chain, where the transverse magnetization follows the same $\tau^{-1/2}$ scaling
for both full and half ramps\cite{puskarov,bialonczyk}

\paragraph{Discussion.}
We have studied near-adiabatic dynamics in the classical transverse field Ising chain. Rich behaviour and scaling are identified based on the ramp type, summarized in Table \ref{table1}. The scaling of physical quantities after the FM to CP ramp follows the Kibble-Zurek prediction.
Most interestingly, the dynamics of the FM to PM ramp encompasses 
a region where the effective dynamics is described
by a non-hermitian Hamiltonian around an exceptional point, emerging from the classical equations of motion  of the parent model.
This effective non-hermitian description incorporates a suppression of the defect production, accounting for the suppressed deviation from adiabaticity 
compared to the half ramp. 

These results are not limited to the one dimensional classical transverse field Ising model, but apply more generally.
By considering $H=\sum_{\langle n,m\rangle } -J S_n^xS_{m}^x-g S_n^z$ for any lattice with uniform coordination in arbitrary dimensions, the very same phase 
diagram and 
effective
Hamiltonian apply not only to the ground state properties but also for the near adiabatic dynamics.

\begin{acknowledgments}
Useful discussions with M. Kormos and G. Tak\'acs are gratefully acknowledged.
This research was supported by the Ministry of Culture and Innovation and the National Research, Development and 
Innovation Office within the Quantum Information National Laboratory of Hungary (Grant No. 2022-2.1.1-NL-2022-00004) 
K134437, K142179  and by a grant of the Ministry of Research, Innovation and Digitization, CNCS/CCCDI-UEFISCDI, under 
projects number PN-III-P4-ID-PCE-2020-0277. This work was in part supported by the Deutsche Forschungsgemeinschaft under grants cluster of excellence ct.qmat (EXC 2147, project-id 390858490). 
\end{acknowledgments}

\bibliographystyle{apsrev}
\bibliography{wboson1}

\begin{thebibliography}{10}
\expandafter\ifx\csname bibnamefont\endcsname\relax
  \def\bibnamefont#1{#1}\fi
\expandafter\ifx\csname bibfnamefont\endcsname\relax
  \def\bibfnamefont#1{#1}\fi
\expandafter\ifx\csname url\endcsname\relax
  \def\url#1{\texttt{#1}}\fi
\expandafter\ifx\csname urlprefix\endcsname\relax\def\urlprefix{URL }\fi
\providecommand{\bibinfo}[2]{#2}
\providecommand{\eprint}[2][]{\url{#2}}

\bibitem{polkovnikovrmp}
\bibinfo{author}{\bibfnamefont{A.}~\bibnamefont{Polkovnikov}},
  \bibinfo{author}{\bibfnamefont{K.}~\bibnamefont{Sengupta}},
  \bibinfo{author}{\bibfnamefont{A.}~\bibnamefont{Silva}}, \bibnamefont{and}
  \bibinfo{author}{\bibfnamefont{M.}~\bibnamefont{Vengalattore}},
  \emph{\bibinfo{title}{\textit{Colloquium} : Nonequilibrium dynamics of closed
  interacting quantum systems}}, \bibinfo{journal}{Rev. Mod. Phys.}
  \textbf{\bibinfo{volume}{83}}, \bibinfo{pages}{863} (\bibinfo{year}{2011}).

\bibitem{dziarmagareview}
\bibinfo{author}{\bibfnamefont{J.}~\bibnamefont{Dziarmaga}},
  \emph{\bibinfo{title}{Dynamics of a quantum phase transition and relaxation
  to a steady state}}, \bibinfo{journal}{Adv. Phys.}
  \textbf{\bibinfo{volume}{59}}, \bibinfo{pages}{1063} (\bibinfo{year}{2010}).

\bibitem{kibble}
\bibinfo{author}{\bibfnamefont{T.~W.~B.} \bibnamefont{Kibble}},
  \emph{\bibinfo{title}{Topology of cosmic domains and strings}},
  \bibinfo{journal}{J. Phys. A} \textbf{\bibinfo{volume}{9}},
  \bibinfo{pages}{1387} (\bibinfo{year}{1976}).

\bibitem{zurek}
\bibinfo{author}{\bibfnamefont{W.~H.} \bibnamefont{Zurek}},
  \emph{\bibinfo{title}{Cosmological experiments in superfluid helium?}},
  \bibinfo{journal}{Nature} \textbf{\bibinfo{volume}{317}},
  \bibinfo{pages}{505} (\bibinfo{year}{1985}).

\bibitem{zurekprl}
\bibinfo{author}{\bibfnamefont{W.~H.} \bibnamefont{Zurek}},
  \bibinfo{author}{\bibfnamefont{U.}~\bibnamefont{Dorner}}, \bibnamefont{and}
  \bibinfo{author}{\bibfnamefont{P.}~\bibnamefont{Zoller}},
  \emph{\bibinfo{title}{Dynamics of a quantum phase transition}},
  \bibinfo{journal}{Phys. Rev. Lett.} \textbf{\bibinfo{volume}{95}},
  \bibinfo{pages}{105701} (\bibinfo{year}{2005}).

\bibitem{dziarmaga}
\bibinfo{author}{\bibfnamefont{J.}~\bibnamefont{Dziarmaga}},
  \emph{\bibinfo{title}{Dynamics of a quantum phase transition: Exact solution
  of the quantum ising model}}, \bibinfo{journal}{Phys. Rev. Lett.}
  \textbf{\bibinfo{volume}{95}}, \bibinfo{pages}{245701}
  (\bibinfo{year}{2005}).

\bibitem{Elliott_TFIM_PRL}
\bibinfo{author}{\bibfnamefont{R.~J.} \bibnamefont{{Elliott}}},
  \bibinfo{author}{\bibfnamefont{P.}~\bibnamefont{{Pfeuty}}}, \bibnamefont{and}
  \bibinfo{author}{\bibfnamefont{C.}~\bibnamefont{{Wood}}},
  \emph{\bibinfo{title}{{Ising Model with a Transverse Field}}},
  \bibinfo{journal}{\prl} \textbf{\bibinfo{volume}{25}}(\bibinfo{number}{7}),
  \bibinfo{pages}{443} (\bibinfo{year}{1970}).

\bibitem{sachdev}
\bibinfo{author}{\bibfnamefont{S.}~\bibnamefont{Sachdev}},
  \emph{\bibinfo{title}{Quantum Phase Transitions}}
  (\bibinfo{publisher}{Cambridge Univ. Press}, \bibinfo{address}{Cambridge},
  \bibinfo{year}{1999}).

\bibitem{mussardo}
\bibinfo{author}{\bibfnamefont{G.}~\bibnamefont{Mussardo}},
  \emph{\bibinfo{title}{{Statistical field theory: an introduction to exactly
  solved models in statistical physics; 1st ed.}}}, Oxford graduate texts
  (\bibinfo{publisher}{Oxford Univ. Press}, \bibinfo{address}{New York, NY},
  \bibinfo{year}{2010}).

\bibitem{laumann}
\bibinfo{author}{\bibfnamefont{C.}~\bibnamefont{Laumann}} \bibnamefont{and}
  \bibinfo{author}{\bibfnamefont{A.}~\bibnamefont{Kitaev}},
  \emph{\bibinfo{title}{Topological phases and quantum computation}},
  \bibinfo{note}{{a}rXiv:0904.2771}.

\bibitem{kitaev2001}
\bibinfo{author}{\bibfnamefont{A.~Y.} \bibnamefont{Kitaev}},
  \emph{\bibinfo{title}{Unpaired majorana fermions in quantum wires}},
  \bibinfo{journal}{Physics-Uspekhi}
  \textbf{\bibinfo{volume}{44}}(\bibinfo{number}{10S}), \bibinfo{pages}{131}
  (\bibinfo{year}{2001}).

\bibitem{bitko}
\bibinfo{author}{\bibfnamefont{D.}~\bibnamefont{Bitko}},
  \bibinfo{author}{\bibfnamefont{T.~F.} \bibnamefont{Rosenbaum}},
  \bibnamefont{and} \bibinfo{author}{\bibfnamefont{G.}~\bibnamefont{Aeppli}},
  \emph{\bibinfo{title}{Quantum critical behavior for a model magnet}},
  \bibinfo{journal}{Phys. Rev. Lett.} \textbf{\bibinfo{volume}{77}},
  \bibinfo{pages}{940} (\bibinfo{year}{1996}).

\bibitem{kinross}
\bibinfo{author}{\bibfnamefont{A.~W.} \bibnamefont{Kinross}},
  \bibinfo{author}{\bibfnamefont{M.}~\bibnamefont{Fu}},
  \bibinfo{author}{\bibfnamefont{T.~J.} \bibnamefont{Munsie}},
  \bibinfo{author}{\bibfnamefont{H.~A.} \bibnamefont{Dabkowska}},
  \bibinfo{author}{\bibfnamefont{G.~M.} \bibnamefont{Luke}},
  \bibinfo{author}{\bibfnamefont{S.}~\bibnamefont{Sachdev}}, \bibnamefont{and}
  \bibinfo{author}{\bibfnamefont{T.}~\bibnamefont{Imai}},
  \emph{\bibinfo{title}{Evolution of quantum fluctuations near the quantum
  critical point of the transverse field ising chain system
  ${\mathrm{conb}}_{2}{\mathrm{o}}_{6}$}}, \bibinfo{journal}{Phys. Rev. X}
  \textbf{\bibinfo{volume}{4}}, \bibinfo{pages}{031008} (\bibinfo{year}{2014}).

\bibitem{keesling}
\bibinfo{author}{\bibfnamefont{A.}~\bibnamefont{Keesling}},
  \bibinfo{author}{\bibfnamefont{A.}~\bibnamefont{Omran}},
  \bibinfo{author}{\bibfnamefont{H.}~\bibnamefont{Levine}},
  \bibinfo{author}{\bibfnamefont{H.}~\bibnamefont{Bernien}},
  \bibinfo{author}{\bibfnamefont{H.}~\bibnamefont{Pichler}},
  \bibinfo{author}{\bibfnamefont{S.}~\bibnamefont{Choi}},
  \bibinfo{author}{\bibfnamefont{R.}~\bibnamefont{Samajdar}},
  \bibinfo{author}{\bibfnamefont{S.}~\bibnamefont{Schwartz}},
  \bibinfo{author}{\bibfnamefont{P.}~\bibnamefont{Silvi}},
  \bibinfo{author}{\bibfnamefont{S.}~\bibnamefont{Sachdev}},
  \bibinfo{author}{\bibfnamefont{P.}~\bibnamefont{Zoller}},
  \bibinfo{author}{\bibfnamefont{M.}~\bibnamefont{Endres}}, \emph{et~al.},
  \emph{\bibinfo{title}{Quantum kibble-zurek mechanism and critical dynamics on
  a programmable rydberg simulator}}, \bibinfo{journal}{Nature}
  \textbf{\bibinfo{volume}{568}}, \bibinfo{pages}{207} (\bibinfo{year}{2019}).

\bibitem{gerling}
\bibinfo{author}{\bibfnamefont{R.~W.} \bibnamefont{Gerling}} \bibnamefont{and}
  \bibinfo{author}{\bibfnamefont{D.~P.} \bibnamefont{Landau}},
  \emph{\bibinfo{title}{Comment on anomalous spin diffusion in classical
  heisenberg magnets.}}, \bibinfo{journal}{Phys. Rev. Lett.}
  \textbf{\bibinfo{volume}{63}}, \bibinfo{pages}{812} (\bibinfo{year}{1989}).

\bibitem{schubert}
\bibinfo{author}{\bibfnamefont{D.}~\bibnamefont{Schubert}},
  \bibinfo{author}{\bibfnamefont{J.}~\bibnamefont{Richter}},
  \bibinfo{author}{\bibfnamefont{F.}~\bibnamefont{Jin}},
  \bibinfo{author}{\bibfnamefont{K.}~\bibnamefont{Michielsen}},
  \bibinfo{author}{\bibfnamefont{H.}~\bibnamefont{De~Raedt}}, \bibnamefont{and}
  \bibinfo{author}{\bibfnamefont{R.}~\bibnamefont{Steinigeweg}},
  \emph{\bibinfo{title}{Quantum versus classical dynamics in spin models:
  Chains, ladders, and square lattices}}, \bibinfo{journal}{Phys. Rev. B}
  \textbf{\bibinfo{volume}{104}}, \bibinfo{pages}{054415}
  (\bibinfo{year}{2021}).

\bibitem{Heisenberg_KPZ}
\bibinfo{author}{\bibfnamefont{A.~J.} \bibnamefont{{McRoberts}}},
  \bibinfo{author}{\bibfnamefont{T.}~\bibnamefont{{Bilitewski}}},
  \bibinfo{author}{\bibfnamefont{M.}~\bibnamefont{{Haque}}}, \bibnamefont{and}
  \bibinfo{author}{\bibfnamefont{R.}~\bibnamefont{{Moessner}}},
  \emph{\bibinfo{title}{{Anomalous dynamics and equilibration in the classical
  Heisenberg chain}}}, \bibinfo{journal}{\prb}
  \textbf{\bibinfo{volume}{105}}(\bibinfo{number}{10}), \bibinfo{eid}{L100403}
  (\bibinfo{year}{2022}).

\bibitem{doyon}
\bibinfo{author}{\bibfnamefont{B.}~\bibnamefont{Doyon}},
  \emph{\bibinfo{title}{{Lecture notes on Generalised Hydrodynamics}}},
  \bibinfo{journal}{SciPost Phys. Lect. Notes} p.~\bibinfo{pages}{18}
  (\bibinfo{year}{2020}).

\bibitem{blundell}
\bibinfo{author}{\bibfnamefont{S.}~\bibnamefont{Blundell}},
  \emph{\bibinfo{title}{Magnetism in Condensed Matter}}, Oxford Master Series
  in Condensed Matter Physics (\bibinfo{publisher}{OUP Oxford},
  \bibinfo{year}{2001}).

\bibitem{nowak}
\bibinfo{author}{\bibfnamefont{U.}~\bibnamefont{Nowak}}, in
  \emph{\bibinfo{booktitle}{Micromagnetism}}, edited by
  \bibinfo{editor}{\bibfnamefont{H.}~\bibnamefont{Kronmüller}},
  \emph{\bibinfo{title}{Classical spin models}} (\bibinfo{publisher}{Wiley},
  \bibinfo{address}{Chichester}, \bibinfo{year}{2007}), no.~\bibinfo{number}{2}
  in \bibinfo{series}{Handbook of magnetism and advanced magnetic materials},
  pp. \bibinfo{pages}{858--876}.

\bibitem{rojas}
\bibinfo{author}{\bibfnamefont{R.}~\bibnamefont{Rojas}},
  \emph{\bibinfo{title}{Neural Networks—A Systematic Introduction}}
  (\bibinfo{publisher}{Springer-Verlag}, \bibinfo{address}{Berlin},
  \bibinfo{year}{1996}).

\bibitem{berloff}
\bibinfo{author}{\bibfnamefont{N.~G.} \bibnamefont{Berloff}},
  \bibinfo{author}{\bibfnamefont{M.}~\bibnamefont{Silva}},
  \bibinfo{author}{\bibfnamefont{K.}~\bibnamefont{Kalinin}},
  \bibinfo{author}{\bibfnamefont{A.}~\bibnamefont{Askitopoulos}},
  \bibinfo{author}{\bibfnamefont{J.~D.} \bibnamefont{T{\"o}pfer}},
  \bibinfo{author}{\bibfnamefont{P.}~\bibnamefont{Cilibrizzi}},
  \bibinfo{author}{\bibfnamefont{W.}~\bibnamefont{Langbein}}, \bibnamefont{and}
  \bibinfo{author}{\bibfnamefont{P.~G.} \bibnamefont{Lagoudakis}},
  \emph{\bibinfo{title}{Realizing the classical xy hamiltonian in polariton
  simulators}}, \bibinfo{journal}{Nature Mater.} \textbf{\bibinfo{volume}{16}},
  \bibinfo{pages}{1120} (\bibinfo{year}{2017}).

\bibitem{parto}
\bibinfo{author}{\bibfnamefont{M.}~\bibnamefont{Parto}},
  \bibinfo{author}{\bibfnamefont{W.}~\bibnamefont{Hayenga}},
  \bibinfo{author}{\bibfnamefont{A.}~\bibnamefont{Marandi}},
  \bibinfo{author}{\bibfnamefont{D.~N.} \bibnamefont{Christodoulides}},
  \bibnamefont{and}
  \bibinfo{author}{\bibfnamefont{M.}~\bibnamefont{Khajavikhan}},
  \emph{\bibinfo{title}{Realizing spin hamiltonians in nanoscale active
  photonic lattices}}, \bibinfo{journal}{Nature Mater.}
  \textbf{\bibinfo{volume}{19}}, \bibinfo{pages}{725} (\bibinfo{year}{2020}).

\bibitem{hoang}
\bibinfo{author}{\bibfnamefont{T.~M.} \bibnamefont{Hoang}},
  \bibinfo{author}{\bibfnamefont{H.~M.} \bibnamefont{Bharath}},
  \bibinfo{author}{\bibfnamefont{M.~J.} \bibnamefont{Boguslawski}},
  \bibinfo{author}{\bibfnamefont{M.}~\bibnamefont{Anquez}},
  \bibinfo{author}{\bibfnamefont{B.~A.} \bibnamefont{Robbins}},
  \bibnamefont{and} \bibinfo{author}{\bibfnamefont{M.~S.}
  \bibnamefont{Chapman}}, \emph{\bibinfo{title}{Adiabatic quenches and
  characterization of amplitude excitations in a continuous quantum phase
  transition}}, \bibinfo{journal}{Proceedings of the National Academy of
  Sciences} \textbf{\bibinfo{volume}{113}}(\bibinfo{number}{34}),
  \bibinfo{pages}{9475} (\bibinfo{year}{2016}).

\bibitem{opatrny}
\bibinfo{author}{\bibfnamefont{T.}~\bibnamefont{Opatrny}},
  \bibinfo{author}{\bibfnamefont{L.}~\bibnamefont{Richterek}},
  \bibnamefont{and} \bibinfo{author}{\bibfnamefont{M.}~\bibnamefont{Opatrny}},
  \emph{\bibinfo{title}{Analogies of the classical euler top with a rotor to
  spin squeezing and quantum phase transitions in a generalized
  lipkin-meshkov-glick model}}, \bibinfo{journal}{Sci. Rep.}
  \textbf{\bibinfo{volume}{8}}, \bibinfo{pages}{1984} (\bibinfo{year}{2018}).

\bibitem{cuccoli}
\bibinfo{author}{\bibfnamefont{A.}~\bibnamefont{Cuccoli}},
  \bibinfo{author}{\bibfnamefont{A.}~\bibnamefont{Taiti}},
  \bibinfo{author}{\bibfnamefont{R.}~\bibnamefont{Vaia}}, \bibnamefont{and}
  \bibinfo{author}{\bibfnamefont{P.}~\bibnamefont{Verrucchi}},
  \emph{\bibinfo{title}{Extracting signatures of quantum criticality in the
  finite-temperature behavior of many-body systems}}, \bibinfo{journal}{Phys.
  Rev. B} \textbf{\bibinfo{volume}{76}}, \bibinfo{pages}{064405}
  (\bibinfo{year}{2007}).

\bibitem{caramico}
\bibinfo{author}{\bibfnamefont{A.}~\bibnamefont{{Caramico D'Auria}}},
  \bibinfo{author}{\bibfnamefont{L.}~\bibnamefont{De~Cesare}},
  \bibinfo{author}{\bibfnamefont{M.~T.} \bibnamefont{Mercaldo}},
  \bibnamefont{and} \bibinfo{author}{\bibfnamefont{I.}~\bibnamefont{Rabuffo}},
  \emph{\bibinfo{title}{Quantum-like criticality for a classical transverse
  ising model in 4–$\epsilon$ dimensions}}, \bibinfo{journal}{Eur. Phys. J.
  B} \textbf{\bibinfo{volume}{77}}, \bibinfo{pages}{419}
  (\bibinfo{year}{2010}).

\bibitem{johann}
\bibinfo{author}{\bibfnamefont{A.}~\bibnamefont{Johann}},
  \emph{\bibinfo{title}{Kink solutions of the classical transverse field ising
  chain}}, \bibinfo{journal}{Eur. Phys. J. B} \textbf{\bibinfo{volume}{25}},
  \bibinfo{pages}{53} (\bibinfo{year}{2002}).

\bibitem{botet}
\bibinfo{author}{\bibfnamefont{R.}~\bibnamefont{Botet}} \bibnamefont{and}
  \bibinfo{author}{\bibfnamefont{R.}~\bibnamefont{Jullien}},
  \emph{\bibinfo{title}{Large-size critical behavior of infinitely coordinated
  systems}}, \bibinfo{journal}{Phys. Rev. B} \textbf{\bibinfo{volume}{28}},
  \bibinfo{pages}{3955} (\bibinfo{year}{1983}).

\bibitem{kolodrubetz}
\bibinfo{author}{\bibfnamefont{M.}~\bibnamefont{Kolodrubetz}},
  \bibinfo{author}{\bibfnamefont{B.~K.} \bibnamefont{Clark}}, \bibnamefont{and}
  \bibinfo{author}{\bibfnamefont{D.~A.} \bibnamefont{Huse}},
  \emph{\bibinfo{title}{Nonequilibrium dynamic critical scaling of the quantum
  ising chain}}, \bibinfo{journal}{Phys. Rev. Lett.}
  \textbf{\bibinfo{volume}{109}}, \bibinfo{pages}{015701}
  (\bibinfo{year}{2012}).

\bibitem{defenu}
\bibinfo{author}{\bibfnamefont{N.}~\bibnamefont{Defenu}},
  \bibinfo{author}{\bibfnamefont{T.}~\bibnamefont{Enss}},
  \bibinfo{author}{\bibfnamefont{M.}~\bibnamefont{Kastner}}, \bibnamefont{and}
  \bibinfo{author}{\bibfnamefont{G.}~\bibnamefont{Morigi}},
  \emph{\bibinfo{title}{Dynamical critical scaling of long-range interacting
  quantum magnets}}, \bibinfo{journal}{Phys. Rev. Lett.}
  \textbf{\bibinfo{volume}{121}}, \bibinfo{pages}{240403}
  (\bibinfo{year}{2018}).

\bibitem{xue2018}
\bibinfo{author}{\bibfnamefont{M.}~\bibnamefont{Xue}},
  \bibinfo{author}{\bibfnamefont{S.}~\bibnamefont{Yin}}, \bibnamefont{and}
  \bibinfo{author}{\bibfnamefont{L.}~\bibnamefont{You}},
  \emph{\bibinfo{title}{Universal driven critical dynamics across a quantum
  phase transition in ferromagnetic spinor atomic bose-einstein condensates}},
  \bibinfo{journal}{Phys. Rev. A} \textbf{\bibinfo{volume}{98}},
  \bibinfo{pages}{013619} (\bibinfo{year}{2018}).

\bibitem{acevedo}
\bibinfo{author}{\bibfnamefont{O.~L.} \bibnamefont{Acevedo}},
  \bibinfo{author}{\bibfnamefont{L.}~\bibnamefont{Quiroga}},
  \bibinfo{author}{\bibfnamefont{F.~J.} \bibnamefont{Rodr\'{\i}guez}},
  \bibnamefont{and} \bibinfo{author}{\bibfnamefont{N.~F.}
  \bibnamefont{Johnson}}, \emph{\bibinfo{title}{New dynamical scaling
  universality for quantum networks across adiabatic quantum phase
  transitions}}, \bibinfo{journal}{Phys. Rev. Lett.}
  \textbf{\bibinfo{volume}{112}}, \bibinfo{pages}{030403}
  (\bibinfo{year}{2014}).

\bibitem{caneva}
\bibinfo{author}{\bibfnamefont{T.}~\bibnamefont{Caneva}},
  \bibinfo{author}{\bibfnamefont{R.}~\bibnamefont{Fazio}}, \bibnamefont{and}
  \bibinfo{author}{\bibfnamefont{G.~E.} \bibnamefont{Santoro}},
  \emph{\bibinfo{title}{Adiabatic quantum dynamics of the lipkin-meshkov-glick
  model}}, \bibinfo{journal}{Phys. Rev. B} \textbf{\bibinfo{volume}{78}},
  \bibinfo{pages}{104426} (\bibinfo{year}{2008}).

\bibitem{Gubser_KZ}
\bibinfo{author}{\bibfnamefont{A.}~\bibnamefont{{Chandran}}},
  \bibinfo{author}{\bibfnamefont{A.}~\bibnamefont{{Erez}}},
  \bibinfo{author}{\bibfnamefont{S.~S.} \bibnamefont{{Gubser}}},
  \bibnamefont{and} \bibinfo{author}{\bibfnamefont{S.~L.}
  \bibnamefont{{Sondhi}}}, \emph{\bibinfo{title}{{Kibble-Zurek problem:
  Universality and the scaling limit}}}, \bibinfo{journal}{\prb}
  \textbf{\bibinfo{volume}{86}}(\bibinfo{number}{6}), \bibinfo{eid}{064304}
  (\bibinfo{year}{2012}).

\bibitem{puebla2015}
\bibinfo{author}{\bibfnamefont{M.-J.} \bibnamefont{Hwang}},
  \bibinfo{author}{\bibfnamefont{R.}~\bibnamefont{Puebla}}, \bibnamefont{and}
  \bibinfo{author}{\bibfnamefont{M.~B.} \bibnamefont{Plenio}},
  \emph{\bibinfo{title}{Quantum phase transition and universal dynamics in the
  rabi model}}, \bibinfo{journal}{Phys. Rev. Lett.}
  \textbf{\bibinfo{volume}{115}}, \bibinfo{pages}{180404}
  (\bibinfo{year}{2015}).

\bibitem{chandranprb}
\bibinfo{author}{\bibfnamefont{A.}~\bibnamefont{Chandran}},
  \bibinfo{author}{\bibfnamefont{A.}~\bibnamefont{Erez}},
  \bibinfo{author}{\bibfnamefont{S.~S.} \bibnamefont{Gubser}},
  \bibnamefont{and} \bibinfo{author}{\bibfnamefont{S.~L.}
  \bibnamefont{Sondhi}}, \emph{\bibinfo{title}{Kibble-zurek problem:
  Universality and the scaling limit}}, \bibinfo{journal}{Phys. Rev. B}
  \textbf{\bibinfo{volume}{86}}, \bibinfo{pages}{064304}
  (\bibinfo{year}{2012}).

\bibitem{adpert}
\bibinfo{author}{\bibfnamefont{C.}~\bibnamefont{De~Grandi}} \bibnamefont{and}
  \bibinfo{author}{\bibfnamefont{A.}~\bibnamefont{Polkovnikov}}, in
  \emph{\bibinfo{booktitle}{Quantum Quenching, Annealing and Computation}},
  edited by \bibinfo{editor}{\bibfnamefont{A.~K.} \bibnamefont{Chandra}},
  \bibinfo{editor}{\bibfnamefont{A.}~\bibnamefont{Das}}, \bibnamefont{and}
  \bibinfo{editor}{\bibfnamefont{B.~K.} \bibnamefont{Chakrabarti}},
  \emph{\bibinfo{title}{Adiabatic perturbation theory: From landau--zener
  problem to quenching through a quantum critical point}}
  (\bibinfo{publisher}{Springer Berlin / Heidelberg}, \bibinfo{year}{2010}),
  vol. \bibinfo{volume}{802} of \emph{\bibinfo{series}{Lecture Notes in
  Physics}}.

\bibitem{cardy}
\bibinfo{author}{\bibfnamefont{J.}~\bibnamefont{Cardy}},
  \emph{\bibinfo{title}{Scaling and Renormalization in Statistical Physics}}
  (\bibinfo{publisher}{Cambridge University Press},
  \bibinfo{address}{Cambridge}, \bibinfo{year}{1996}).

\bibitem{gao2015}
\bibinfo{author}{\bibfnamefont{T.}~\bibnamefont{Gao}},
  \bibinfo{author}{\bibfnamefont{E.}~\bibnamefont{Estrecho}},
  \bibinfo{author}{\bibfnamefont{K.~Y.} \bibnamefont{Bliokh}},
  \bibinfo{author}{\bibfnamefont{T.~C.~H.} \bibnamefont{Liew}},
  \bibinfo{author}{\bibfnamefont{M.~D.} \bibnamefont{Fraser}},
  \bibinfo{author}{\bibfnamefont{S.}~\bibnamefont{Brodbeck}},
  \bibinfo{author}{\bibfnamefont{M.}~\bibnamefont{Kamp}},
  \bibinfo{author}{\bibfnamefont{C.}~\bibnamefont{Schneider}},
  \bibinfo{author}{\bibfnamefont{S.}~\bibnamefont{H{\"o}fling}},
  \bibinfo{author}{\bibfnamefont{Y.}~\bibnamefont{Yamamoto}},
  \bibinfo{author}{\bibfnamefont{F.}~\bibnamefont{Nori}},
  \bibinfo{author}{\bibfnamefont{Y.~S.} \bibnamefont{Kivshar}}, \emph{et~al.},
  \emph{\bibinfo{title}{Observation of non-hermitian degeneracies in a chaotic
  exciton-polariton billiard}}, \bibinfo{journal}{Nature}
  \textbf{\bibinfo{volume}{526}}, \bibinfo{pages}{554} (\bibinfo{year}{2015}).

\bibitem{rotter}
\bibinfo{author}{\bibfnamefont{I.}~\bibnamefont{Rotter}} \bibnamefont{and}
  \bibinfo{author}{\bibfnamefont{J.~P.} \bibnamefont{Bird}},
  \emph{\bibinfo{title}{A review of progress in the physics of open quantum
  systems: theory and experiment}}, \bibinfo{journal}{Rep. Prog. Phys.}
  \textbf{\bibinfo{volume}{78}}, \bibinfo{pages}{114001}
  (\bibinfo{year}{2015}).

\bibitem{zeuner}
\bibinfo{author}{\bibfnamefont{J.~M.} \bibnamefont{Zeuner}},
  \bibinfo{author}{\bibfnamefont{M.~C.} \bibnamefont{Rechtsman}},
  \bibinfo{author}{\bibfnamefont{Y.}~\bibnamefont{Plotnik}},
  \bibinfo{author}{\bibfnamefont{Y.}~\bibnamefont{Lumer}},
  \bibinfo{author}{\bibfnamefont{S.}~\bibnamefont{Nolte}},
  \bibinfo{author}{\bibfnamefont{M.~S.} \bibnamefont{Rudner}},
  \bibinfo{author}{\bibfnamefont{M.}~\bibnamefont{Segev}}, \bibnamefont{and}
  \bibinfo{author}{\bibfnamefont{A.}~\bibnamefont{Szameit}},
  \emph{\bibinfo{title}{Observation of a topological transition in the bulk of
  a non-hermitian system}}, \bibinfo{journal}{Phys. Rev. Lett.}
  \textbf{\bibinfo{volume}{115}}, \bibinfo{pages}{040402}
  (\bibinfo{year}{2015}).

\bibitem{Feng2014}
\bibinfo{author}{\bibfnamefont{L.}~\bibnamefont{Feng}},
  \bibinfo{author}{\bibfnamefont{Z.~J.} \bibnamefont{Wong}},
  \bibinfo{author}{\bibfnamefont{R.-M.} \bibnamefont{Ma}},
  \bibinfo{author}{\bibfnamefont{Y.}~\bibnamefont{Wang}}, \bibnamefont{and}
  \bibinfo{author}{\bibfnamefont{X.}~\bibnamefont{Zhang}},
  \emph{\bibinfo{title}{Single-mode laser by parity-time symmetry breaking}},
  \bibinfo{journal}{Science}
  \textbf{\bibinfo{volume}{346}}(\bibinfo{number}{6212}), \bibinfo{pages}{972}
  (\bibinfo{year}{2014}).

\bibitem{hodaei}
\bibinfo{author}{\bibfnamefont{H.}~\bibnamefont{Hodaei}},
  \bibinfo{author}{\bibfnamefont{A.~U.} \bibnamefont{Hassan}},
  \bibinfo{author}{\bibfnamefont{S.}~\bibnamefont{Wittek}},
  \bibinfo{author}{\bibfnamefont{H.}~\bibnamefont{Garcia-Gracia}},
  \bibinfo{author}{\bibfnamefont{R.}~\bibnamefont{El-Ganainy}},
  \bibinfo{author}{\bibfnamefont{D.~N.} \bibnamefont{Christodoulides}},
  \bibnamefont{and}
  \bibinfo{author}{\bibfnamefont{M.}~\bibnamefont{Khajavikhan}},
  \emph{\bibinfo{title}{Enhanced sensitivity at higher-order exceptional
  points}}, \bibinfo{journal}{Nature} \textbf{\bibinfo{volume}{548}},
  \bibinfo{pages}{187} (\bibinfo{year}{2017}).

\bibitem{Bergholtz2021}
\bibinfo{author}{\bibfnamefont{E.~J.} \bibnamefont{Bergholtz}},
  \bibinfo{author}{\bibfnamefont{J.~C.} \bibnamefont{Budich}},
  \bibnamefont{and} \bibinfo{author}{\bibfnamefont{F.~K.} \bibnamefont{Kunst}},
  \emph{\bibinfo{title}{Exceptional topology of non-hermitian systems}},
  \bibinfo{journal}{Rev. Mod. Phys.} \textbf{\bibinfo{volume}{93}},
  \bibinfo{pages}{015005} (\bibinfo{year}{2021}).

\bibitem{ashidareview}
\bibinfo{author}{\bibfnamefont{Y.}~\bibnamefont{Ashida}},
  \bibinfo{author}{\bibfnamefont{Z.}~\bibnamefont{Gong}}, \bibnamefont{and}
  \bibinfo{author}{\bibfnamefont{M.}~\bibnamefont{Ueda}},
  \emph{\bibinfo{title}{Non-hermitian physics}}, \bibinfo{journal}{Advances in
  Physics} \textbf{\bibinfo{volume}{69}}, \bibinfo{pages}{3}
  (\bibinfo{year}{2020}).

\bibitem{ElGanainy2018}
\bibinfo{author}{\bibfnamefont{R.}~\bibnamefont{El-Ganainy}},
  \bibinfo{author}{\bibfnamefont{K.~G.} \bibnamefont{Makris}},
  \bibinfo{author}{\bibfnamefont{M.}~\bibnamefont{Khajavikhan}},
  \bibinfo{author}{\bibfnamefont{Z.~H.} \bibnamefont{Musslimani}},
  \bibinfo{author}{\bibfnamefont{S.}~\bibnamefont{Rotter}}, \bibnamefont{and}
  \bibinfo{author}{\bibfnamefont{D.~N.} \bibnamefont{Christodoulides}},
  \emph{\bibinfo{title}{Non-hermitian physics and pt symmetry}},
  \bibinfo{journal}{Nat. Phys.}
  \textbf{\bibinfo{volume}{14}}(\bibinfo{number}{1}), \bibinfo{pages}{11}
  (\bibinfo{year}{2018}).

\bibitem{fruchart}
\bibinfo{author}{\bibfnamefont{M.}~\bibnamefont{Fruchart}},
  \bibinfo{author}{\bibfnamefont{R.}~\bibnamefont{Hanai}},
  \bibinfo{author}{\bibfnamefont{P.~B.} \bibnamefont{Littlewood}},
  \bibnamefont{and} \bibinfo{author}{\bibfnamefont{V.}~\bibnamefont{Vitelli}},
  \emph{\bibinfo{title}{Non-reciprocal phase transitions}},
  \bibinfo{journal}{Nature} \textbf{\bibinfo{volume}{592}},
  \bibinfo{pages}{363} (\bibinfo{year}{2021}).

\bibitem{boozer}
\bibinfo{author}{\bibfnamefont{A.~D.} \bibnamefont{Boozer}},
  \emph{\bibinfo{title}{Quantum field theory in (0 + 1) dimensions}},
  \bibinfo{journal}{European Journal of Physics}
  \textbf{\bibinfo{volume}{28}}(\bibinfo{number}{4}), \bibinfo{pages}{729}
  (\bibinfo{year}{2007}).

\bibitem{mostafazadeh2002}
\bibinfo{author}{\bibfnamefont{A.}~\bibnamefont{Mostafazadeh}},
  \emph{\bibinfo{title}{Pseudo-hermiticity versus pt symmetry: The necessary
  condition for the reality of the spectrum of a non-hermitian hamiltonian}},
  \bibinfo{journal}{Journal of Mathematical Physics}
  \textbf{\bibinfo{volume}{43}}(\bibinfo{number}{1}), \bibinfo{pages}{205}
  (\bibinfo{year}{2002}).

\bibitem{mostafazadeh2003}
\bibinfo{author}{\bibfnamefont{A.}~\bibnamefont{Mostafazadeh}},
  \emph{\bibinfo{title}{{Exact PT}-symmetry is equivalent to hermiticity}},
  \bibinfo{journal}{Journal of Physics A: Mathematical and General}
  \textbf{\bibinfo{volume}{36}}(\bibinfo{number}{25}), \bibinfo{pages}{7081}
  (\bibinfo{year}{2003}).

\bibitem{Bender2007}
\bibinfo{author}{\bibfnamefont{C.~M.} \bibnamefont{Bender}},
  \emph{\bibinfo{title}{Making sense of non-hermitian hamiltonians}},
  \bibinfo{journal}{Reports on Progress in Physics}
  \textbf{\bibinfo{volume}{70}}(\bibinfo{number}{6}), \bibinfo{pages}{947}
  (\bibinfo{year}{2007}).

\bibitem{heiss}
\bibinfo{author}{\bibfnamefont{W.~D.} \bibnamefont{Heiss}},
  \emph{\bibinfo{title}{The physics of exceptional points}},
  \bibinfo{journal}{J. Phys. A: Math. Theor.}
  \textbf{\bibinfo{volume}{45}}(\bibinfo{number}{44}), \bibinfo{pages}{444016}
  (\bibinfo{year}{2012}).

\bibitem{Note1}
\bibinfo{note}{In principle, the initial condition is $\delta ^x(\tau /g)\sim
  \tau ^{-1/3}$ and $\delta ^y(\tau /g)\sim \tau ^{-2/3}\rightarrow 0$ where
  the latter is taken to zero as it is parametrically smaller than the former
  in the adiabatic, $\tau \rightarrow \infty $ limit.}

\bibitem{dorakz}
\bibinfo{author}{\bibfnamefont{B.}~\bibnamefont{D{\'o}ra}},
  \bibinfo{author}{\bibfnamefont{M.}~\bibnamefont{Heyl}}, \bibnamefont{and}
  \bibinfo{author}{\bibfnamefont{R.}~\bibnamefont{Moessner}},
  \emph{\bibinfo{title}{The kibble-zurek mechanism at exceptional points}},
  \bibinfo{journal}{Nature Communications}
  \textbf{\bibinfo{volume}{10}}(\bibinfo{number}{1}), \bibinfo{pages}{2254}
  (\bibinfo{year}{2019}).

\bibitem{PRXQ}
\bibinfo{author}{\bibfnamefont{L.}~\bibnamefont{Xiao}},
  \bibinfo{author}{\bibfnamefont{D.}~\bibnamefont{Qu}},
  \bibinfo{author}{\bibfnamefont{K.}~\bibnamefont{Wang}},
  \bibinfo{author}{\bibfnamefont{H.-W.} \bibnamefont{Li}},
  \bibinfo{author}{\bibfnamefont{J.-Y.} \bibnamefont{Dai}},
  \bibinfo{author}{\bibfnamefont{B.}~\bibnamefont{D\'ora}},
  \bibinfo{author}{\bibfnamefont{M.}~\bibnamefont{Heyl}},
  \bibinfo{author}{\bibfnamefont{R.}~\bibnamefont{Moessner}},
  \bibinfo{author}{\bibfnamefont{W.}~\bibnamefont{Yi}}, \bibnamefont{and}
  \bibinfo{author}{\bibfnamefont{P.}~\bibnamefont{Xue}},
  \emph{\bibinfo{title}{Non-hermitian kibble-zurek mechanism with tunable
  complexity in single-photon interferometry}}, \bibinfo{journal}{PRX Quantum}
  \textbf{\bibinfo{volume}{2}}, \bibinfo{pages}{020313} (\bibinfo{year}{2021}).

\bibitem{graefe2008}
\bibinfo{author}{\bibfnamefont{E.~M.} \bibnamefont{Graefe}},
  \bibinfo{author}{\bibfnamefont{H.~J.} \bibnamefont{Korsch}},
  \bibnamefont{and} \bibinfo{author}{\bibfnamefont{A.~E.}
  \bibnamefont{Niederle}}, \emph{\bibinfo{title}{Mean-field dynamics of a
  non-hermitian bose-hubbard dimer}}, \bibinfo{journal}{Phys. Rev. Lett.}
  \textbf{\bibinfo{volume}{101}}, \bibinfo{pages}{150408}
  (\bibinfo{year}{2008}).

\bibitem{orden}
\bibinfo{author}{\bibfnamefont{J.~W.~V.} \bibnamefont{Orden}},
  \bibinfo{author}{\bibfnamefont{S.}~\bibnamefont{Jeschonnek}},
  \bibnamefont{and} \bibinfo{author}{\bibfnamefont{J.}~\bibnamefont{Tjon}},
  \emph{\bibinfo{title}{Scaling of dirac fermions and the wkb approximation}},
  \bibinfo{journal}{Phys. Rev. D} \textbf{\bibinfo{volume}{72}},
  \bibinfo{pages}{054020} (\bibinfo{year}{2005}).

\bibitem{fei}
\bibinfo{author}{\bibfnamefont{Z.}~\bibnamefont{Fei}},
  \bibinfo{author}{\bibfnamefont{N.}~\bibnamefont{Freitas}},
  \bibinfo{author}{\bibfnamefont{V.}~\bibnamefont{Cavina}},
  \bibinfo{author}{\bibfnamefont{H.~T.} \bibnamefont{Quan}}, \bibnamefont{and}
  \bibinfo{author}{\bibfnamefont{M.}~\bibnamefont{Esposito}},
  \emph{\bibinfo{title}{Work statistics across a quantum phase transition}},
  \bibinfo{journal}{Phys. Rev. Lett.} \textbf{\bibinfo{volume}{124}},
  \bibinfo{pages}{170603} (\bibinfo{year}{2020}).

\bibitem{puskarov}
\bibinfo{author}{\bibfnamefont{T.}~\bibnamefont{Puskarov}} \bibnamefont{and}
  \bibinfo{author}{\bibfnamefont{D.}~\bibnamefont{Schuricht}},
  \emph{\bibinfo{title}{{Time evolution during and after finite-time quantum
  quenches in the transverse-field Ising chain}}}, \bibinfo{journal}{SciPost
  Phys.} \textbf{\bibinfo{volume}{1}}, \bibinfo{pages}{003}
  (\bibinfo{year}{2016}).

\bibitem{bialonczyk}
\bibinfo{author}{\bibfnamefont{M.}~\bibnamefont{Bialonczyk}} \bibnamefont{and}
  \bibinfo{author}{\bibfnamefont{B.}~\bibnamefont{Damski}},
  \emph{\bibinfo{title}{One-half of the kibble–zurek quench followed by free
  evolution}}, \bibinfo{journal}{Journal of Statistical Mechanics: Theory and
  Experiment} \textbf{\bibinfo{volume}{2018}}(\bibinfo{number}{7}),
  \bibinfo{pages}{073105} (\bibinfo{year}{2018}).

\end{thebibliography}

\end{document}